\documentclass[twocolumn,showpacs,preprintnumbers,amsmath,amssymb]{revtex4}

\usepackage{graphicx}
\usepackage{dcolumn}
\usepackage{bm}

\begin{document}

\preprint{APS/123-QED}

\title{Trading interactions for topology in scale-free networks}

\author{C.V. Giuraniuc$^{1}$, J.P.L. Hatchett$^{2}$, J.O. Indekeu$^{1}$, M.
Leone$^{3}$, I. P\'erez Castillo$^{4}$, B. Van Schaeybroeck$^{1}$,
C. Vanderzande $^{4,5}$}
\affiliation{%
$^{1}$Laboratorium voor Vaste-Stoffysica en Magnetisme, Katholieke
Universiteit Leuven, 3001 Leuven, Belgium,\\$^{2}$Department of
Mathematics, King's College London, The Strand, London WC2R 2LS,
UK,
\\$^{3}$Institute for Scientific Interchange (I.S.I.),
Villa Gualino, Viale Settimio Severo 65, 10133 Turin, Italy,\\
$^{4}$Institute for Theoretical Physics, Katholieke Universiteit
Leuven, 3001 Leuven, Belgium,\\$^{5}$Departement WNI, Hasselt
University, 3590 Diepenbeek, Belgium
}%

\date{\today}

\begin{abstract}
Scale-free networks with topology-dependent interactions are
studied. It is shown that the universality classes of critical
behavior, which conventionally depend only on topology, can also
be explored by tuning the interactions. A mapping, $\gamma' =
(\gamma - \mu)/(1-\mu)$, describes how a shift of the standard
exponent $\gamma$ of the degree distribution $P(q)$ can absorb the
effect of degree-dependent pair interactions $J_{ij} \propto
(q_iq_j)^{-\mu}$. Replica technique, cavity method and Monte Carlo
simulation support the physical picture suggested by Landau theory
for the critical exponents and by the Bethe-Peierls approximation
for the critical temperature. The equivalence of topology and
interaction holds for equilibrium and non-equilibrium systems, and
is illustrated with interdisciplinary applications.

\end{abstract}

\pacs{89.75.Hc, 64.60.Fr, 05.70.Jk, 05.50.+q}
\maketitle

In this Letter we pose and answer the following fundamental
questions. What are the relevant variables that determine the
universality classes of critical behavior in networks? Is the
conventional classification based on the exponent $\gamma$ of the
distribution function of network connections $P(q)$ sufficient?
What is the effect on the critical behavior of interactions that
depend on the connectivity? Do the interactions give rise to new
relevant variables or can these be ``transformed away"
topologically? We focus on scaling arguments and analytic
approaches \cite{Giu}.

Nowadays, scale-free networks enjoy a lot of attention due to
their ubiquitous occurrence and the suitability of modern
computational techniques for understanding their properties and
predicting their behavior \cite{reviews}. These networks typically
have simple topology. The usual notions of spatial coordinates and
dimensionality are not present; e.g., while the ``volume" scales
as the number of nodes $N$, the largest distance or ``diameter"
grows not faster than $\log N$, suggesting infinite ``dimension".
This is why mean-field or Landau theories are successful.

Understanding critical phenomena in scale-free networks is of
fundamental and practical relevance. There is a surprising
diversity of ``mean-field" universality classes, which are
topology-dependent \cite{Goltsev}. However, for most real networks
and for the simplest interactions between nodes, the critical
point is inaccessible (e.g., $T_c = \infty$). We propose a way
around this by tuning the interactions.

Without loss of generality, consider first the context of opinion
formation in ``sociophysics"\cite{socio}. The network consists of
people. Opinions are represented by spin orientations, and
communication by pair interactions. The statistical physics of the
formation of a common opinion is akin to that of spontaneous
symmetry breaking below a critical temperature $T_c$ in a
Hamiltonian model for equilibrium cooperative phenomena
\cite{Alek}.

Putting Ising (or Potts) spins $s_i$ on the nodes $i=1,..., N$ and
ferromagnetic interactions $J$ on the edges, leads to interesting
critical behavior, different from its counterpart in lattice spin
models \cite{critical,Goltsev}. For the standard scale-free
distribution $P(q) \propto q^{-\gamma}$ of the number of
connections or ``degree" of a node, the decay exponent $\gamma$ is
the key parameter distinguishing the universality classes.
Standard mean-field critical behavior ($\alpha = 0$ for the
specific heat, $\beta = 1/2$ for the order parameter,
$\gamma_{sus} =1$ for the susceptibility) is predicted for $\gamma
> 5$. Logarithmic corrections appear at $\gamma = 5$. For $3 <
\gamma < 5$ the critical exponents $\alpha$ and $\beta$ are
non-universal; they depend on $\gamma$. The susceptibility
exponent $\gamma_{sus} = 1$ is superuniversal. Conventional
finite-size scaling (as a function of network size $N$) predicts
the familiar constraint $\alpha + 2\beta +\gamma_{sus} =2$. For
$\gamma \leq 3$, however, the framework of finite-temperature
critical phenomena breaks down as $T_c$ moves to infinity.

Which values of $\gamma$ apply to real scale-free networks? Most
studied ones, including WWW, Internet, collaboration, citation,
cellular, ecological or linguistic networks have $2 \leq \gamma
\leq 3$. The Barab\'asi-Albert (BA) network, grown by preferential
attachment, has $\gamma =3$.

The initial motivation for introducing degree-dependent
interactions $J_{ij} \equiv J(q_i,q_j)$ was to prevent $T_c$ from
diverging in the Ising model on the BA network \cite{Alek} by
compensating high connectivity with weak interaction through
$J_{ij} \propto 1/\sqrt{q_iq_j}$ \cite{SAN}. Further work has led
us to observe that this interaction is equivalent to a
$q$-independent $J$, provided $\gamma$ is shifted from 3 to 5. A
Landau-theoretic argument and a mean-field {\em ansatz} generalize
this equivalence (see further) and provide the insight that
interactions that depend on connectivity effectively modify the
topological distribution $P(q)$.

We introduce a family of interactions parameterized by an exponent
$\mu$,
\begin{equation}
J_{ij} = J \langle q\rangle ^{2\mu} (q_iq_j)^{-\mu},
\end{equation}
where $\langle .\rangle $ denotes the average over the degree
distribution $P(q)$. Generalizing the Landau theory of Goltsev
{\em et al.} \cite{Goltsev} we obtain the following form, in zero
external field, for the ``free energy" $\Phi$ as a function of the
order parameter $x \equiv \sum_i E_T(s_i)/N$, with $E_T(.)$ the
thermal average,
\begin{equation}
\Phi(x) = \langle \phi(q^{-\mu}x, q^{1-\mu}x)\rangle,
\end{equation}
where the first argument of the constrained free energy $\phi$ is
the rescaled order parameter, and the second one is the rescaled
effective field acting on a node with $q$ connections. Both
rescalings, by a factor $q^{-\mu}$, come from the $q_i$- and
$q_j$-dependence of $J_{ij}$. The crucial assumption
\cite{Goltsev} is that $\phi(y,z)$ can be written as a power
series in $y$ and $z$. The singularity structure of $\Phi(x)$,
leading to fascinating deviations from standard mean-field
behavior, is then induced by the fact that the moments $\langle
q^n\rangle$ diverge for $n \geq n_c = \gamma - 1$. This in turn
leads to a divergence of the coefficient $f_n$ of $x^n$ in the
free energy. Thus, in this system non-classical critical behavior
is possible, not due to spatial fluctuations of the order
parameter (since space is effectively infinite-dimensional), but
due to the scale-free character of the degree distribution.

Our main result is that a network with exponents $(\gamma,\mu)$
can be mapped onto one with $(\gamma',\mu' = 0)$, in the sense
that both are in the same universality class of critical behavior.
The latter has constant couplings, independent of $q$. We find the
exponent relation
\begin{equation}
\gamma' = (\gamma - \mu)/(1-\mu),
\end{equation}
which can be proven as follows. The critical value $n=n_c$ for
diverging coefficients $f_n$ in the free energy must be the same
in order for the two models to be equivalent. Since the leading
moment $\langle q^n\rangle$ in $f_n$ in the model with $\mu=0$
gets replaced by $\langle q^{(1-\mu)n}\rangle $ in the model with
$\mu \neq 0$, we obtain $n_c= \gamma' -1 = (\gamma -1)/(1 - \mu)$.

A more general ``mean-field" proof, which does not rely on a free
energy, runs as follows. Within a mean-field approach the only way
in which the degree $q$ enters in physical properties is through
the {\em quenched average interaction}, over all networks, between
any two nodes with {\em fixed} degrees $q_i$ and $q_j$. This
average is $J_{ij}p_{ij}$, with $p_{ij} = q_i q_j/(\langle
q\rangle N)$ the probability that $i$ and $j$ are connected. Note
that even for constant $J$ (case $\mu = 0$) the quenched average
interaction is $q$-dependent, while for $\mu = 1$ it is not. For
$\mu \neq 0$, the $q_i$ are transformed to $q'_i= q_i^{(1-\mu)}$,
using (1). In order to retain the same physics, averages over the
degree distribution must be invariant. This requires a
distribution transformation,
\begin{equation}
P(q) = P'(q'(q)) dq'(q)/dq,
\end{equation}
from which (3) follows for scale-free (power-law) $P(q)$.

Consequently, for a scale-free network the range $\mu \in
[2-\gamma,1]$ allows one to explore the whole range of
universality classes uncovered in previous works. It is no longer
necessary to vary the network topology. It suffices, for a fixed
$\gamma$, to tune the form of the interaction. An important
consequence is that for real scale-free networks, with typically
$\gamma \leq 3$, one is no longer set back by an infinite critical
temperature when putting on interactions. For the BA network the
limit $\mu \rightarrow 1$ maps onto $\gamma' \rightarrow \infty$
and corresponds to the cross-over to a thin-tailed degree
distribution of, e.g., Poisson type. In the opposite limit, $\mu
\rightarrow -1$, for which $\gamma' \rightarrow 2$, the threshold
is reached beyond which the first moment $\langle q\rangle$
diverges in the equivalent network with constant couplings.

\begin{figure}[b]
\includegraphics[width=2.5in,height=1.8in]{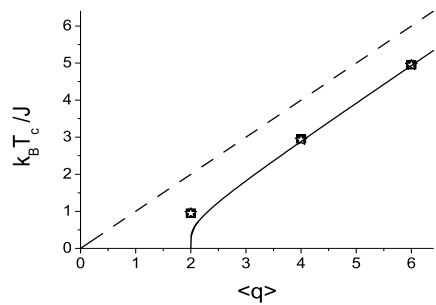}
\caption{\label{fig:epsart} $T_c$ versus $\langle q\rangle$ for a
scale-free network with $\gamma = 3$ and $\mu = 1/2$. BP estimates
(filled squares) almost coincide with exact results (replica
technique, open stars). The solid line is the lattice BP
approximation for $Q = \langle q\rangle$ and the dashed line is
the MF approximation.}
\end{figure}

For locating the critical temperature a Bethe-Peierls (BP)
approximation, which emphasizes the local Cayley tree-like
structure of the network, normally gives a very reasonable first
approximation \cite{critical}. The entropic contribution to the
free energy is truncated to single-spin and pair terms \cite{Mar}.
Besides this standard assumption, we impose a new scaling relation
on the local order parameter, $E_T(s_i)$, in a given network. This
scaling is in harmony with the definition of the effective field
acting on $s_i$ in the Landau theory and is corroborated by MCS.
It reads
\begin{equation}
E_T(s_i) \approx q_i^{1-\mu}\;x/\langle q^{1-\mu}\rangle.
\end{equation}
Using this to eliminate $s_i$ in terms of $q_i$ and $x$ one
obtains the self-consistent BP equation near $T_c$,
\begin{eqnarray}
\langle q\rangle\sum_q P(q)(q-1) q^{1-\mu} = \nonumber\\
\sum_{q_1}P(q_1)q_1^{2-\mu}\sum_{q_2}P(q_2)q_2 \left (
1+\tanh\frac{J \langle q\rangle ^{2\mu}}{k_B T
q_1^{\mu}q_2^{\mu}}\right )^{-1}.
\end{eqnarray}
For the conventional case $\mu = 0$, we can extract from this the
large-$\langle q\rangle$ approximation $k_BT_c/J \approx \langle
q^2\rangle /\langle q\rangle -1$, in agreement with exact results
\cite{critical}.

We now turn to illustrations of this framework and start with
$\gamma = 3$ and $\mu = 1/2$ (``Special Attention Network"
\cite{SAN}). We discuss the critical point, specific heat, order
parameter and susceptibility.

The critical temperature versus average degree $\langle q\rangle$
is shown in Fig.1, for $\langle q\rangle$ = 2, 4, and 6. It has
been derived by exact solution of the model using the replica
technique. Since all couplings are ``ferromagnetic" the replica
symmetry is not broken. For comparison, the almost coincident
results from the BP approximation (6) are also indicated. A good
rule of thumb is $k_BT_c/J \approx \langle q\rangle -1$.
Incidentally, this is also the result of the large-$\langle
q\rangle $ expansion which can be derived from (6). Interestingly,
the results are very close to the conventional BP approximation
$J/k_BT_c = 0.5 \ln [(Q-2)/Q]$ (thin solid line), for a regular
lattice with coordination number $Q$, except for $Q=2$.  For
completeness the mean-field conjecture \cite{SAN} $k_BT_c/J =
\langle q\rangle$ is also shown.

These calculations assume no correlations exist between the edges
of the network. The rule of preferential attachment violates this
assumption, so the BA network must be considered separately. For
the special case $\langle q\rangle =2$, a BA network differs
strongly from a correlation-free network. The former consists of a
single tree-like structure {\em without} loops, whence $T_c = 0$.
In contrast, an uncorrelated network with $\langle q\rangle =2$
can consist of clusters, which can have loops, leading to a finite
$T_c$, as replica technique and BP approximation predict.

Having located the critical temperature, and elucidated its
non-universality through its dependence on $\langle q\rangle$ and
on the network correlations, we now turn to more universal
properties. The specific heat singularity for the Ising model on
scale-free networks has a very interesting and subtle form
\cite{critical}. For $\mu = 0$, it varies from a classic
mean-field jump, predicted for $\gamma > 5$, to a continuous
behavior but with diverging slope, for $4<\gamma<5$, and with
continuous slope for $3<\gamma<4$. For $\mu = 1/2$, the
equivalence relation, Eq.(3), predicts, with $\tau = (T-T_c)/T_c$,
\begin{equation}
C_{sing} \propto (\ln \tau^{-1})^{-1}.
\end{equation}
The shape of this specific heat near $T_c$ is astonishingly
different from that of the conventional mean-field jump. Although
the critical exponent $\alpha$ consistent with (7) is zero, as for
standard mean field, the inverse logarithm ensures a {\em
vanishing} jump, with diverging slope.

Evidence gathered from MCS, and, more convincingly, the cavity
method, shows that i) the specific heat $C(T)$ reaches a maximum
well {\em below} $T_c$ and ii) the jump singularity that is
apparent for small network size $N$ closes slowly when $N$ is
increased. These findings are consistent with (3), and suggest
that for $\gamma = 3$ and $\mu = 1/2$, the network indeed maps
onto one with $\gamma = 5$ and $\mu' = 0$.

Figures 2 and 3 illustrate these results for the specific heat,
for a BA network with $\mu = 1/2$. Shown are MCS (Fig.2) for
$\langle q\rangle =10$ and $N=5600$, and the cavity method (Fig.3)
for $\langle q\rangle =4$, applied to the sequence $N= 100, ...,
10^6$. The latter technique clearly suggests the slow closing of
the specific heat jump, for $N \rightarrow \infty$.

\begin{figure}
\includegraphics[width=2.5in,height=1.8in]{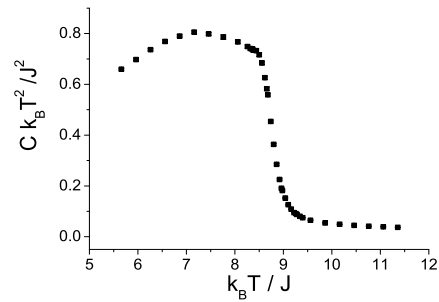}
\caption{\label{fig:epsart} Specific heat from MCS of BA networks
($\gamma = 3$), with $\mu = 1/2$ and $\langle q\rangle =10$. The
critical temperature in the large-$N$ limit is $k_B T_c/J \approx
8.96$ (Eq.6). For $N=5600$, MCS of the susceptibility maximum
indicates $k_B T_c(N)/J \approx 8.65$, well above the temperature
of the specific heat maximum.}
\end{figure}

\begin{figure}
\includegraphics[width=2.5in,height=2in]{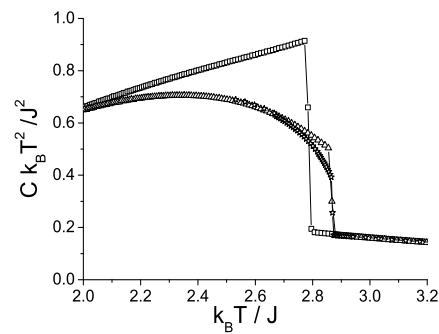}
\caption{\label{fig:epsart} Specific heat from the cavity method
for BA networks, with $\mu = 1/2$ and $\langle q\rangle =4$. Shown
are data for $N=10^2$ (squares), $N=10^4$ (triangles), and
$N=10^6$ (stars). Lines are guides to the eye. The finite-size
critical temperature (apparent from the jump in $C$) rapidly
converges for large $N$, to $k_B T_c/J \approx 2.92$ (Eq.6).
Clearly, the maximum in $C$ precedes $T_c$ and the jump in $C$
tends to close for large $N$.}
\end{figure}

We have also computed the order parameter (``magnetization") and
susceptibility singularities near $T_c$ and find, for $\mu = 1/2$,
that the results are consistent with $\beta = 1/2$ and
$\gamma_{sus} = 1$. Again, these values agree with the predictions
from the mapping (3). It should be remarked that for $\gamma = 5$
and $\mu = 0$ a logarithmic correction factor is predicted for the
order parameter, which, however, cannot easily be detected on top
of the square-root singularity $\beta = 1/2$. Further, our result
for $\gamma_{sus}$ is not discriminative, since $\gamma_{sus} = 1$
is superuniversal, valid for all $\gamma > 3$ in the model with
$\mu = 0$.

\begin{figure}
\includegraphics[width=2.5in,height=2in]{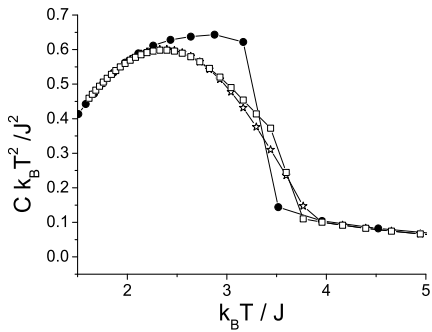}
\caption{\label{fig:epsart} Specific heat from the cavity method
for BA networks, with $\mu = 1/3$ and $\langle q\rangle =4$, for
$N=10^2$ (circles), $N=10^3$ (squares), and $N=10^5$ (stars). A
linear $T$-dependence develops for large $N$ and meets the
high-$T$ background with a jump in slope at $T_c$. The finite-size
critical temperature converges to $k_B T_c/J \approx 4.21$
(Eq.6).}
\end{figure}

Further evidence for (3) is obtained for the interesting case
$\gamma = 3$ and $\mu = 1/3$. A linear specific heat ($\alpha =
-1$) and a linear order parameter singularity ($\beta = 1$)
clearly emerge for large $N$ in cavity method computations
(Fig.4). MCS of specific heat (Fig.5), order parameter and
susceptibility supports this conclusion. The results agree with
what is expected for $\gamma'=4$ and $\mu' = 0$.

Additional evidence for (3) comes from a study of networks with
$\gamma =3$ and $\mu = 1$. Using MCS and cavity method the
standard mean-field jump of the specific heat is retrieved. This
is consistent with a shift of $\gamma$ to a value greater than 5
in the reference system with $\mu = 0$.

We now leave equilibrium statistical mechanics and focus on
dynamical systems. In the contact process for disease spreading
each node can be either ill or healthy. An ill node can cure at
rate 1, and a healthy node becomes infected at a rate which is
$\lambda$ times the number of ill neighbors. The model has a phase
transition between an absorbing healthy state and an active state
with a non-zero density of ill nodes, at some $\lambda_c$. On a
scale-free network a finite $\lambda_c$ is found for $\gamma
> 3$. The critical exponents depend on $\gamma$ in the range $3 <
\gamma \leq 4$ and assume standard mean-field values for $\gamma
> 4$ \cite{contact}. Generalizing the process to a degree-dependent
infection rate, $ \lambda_{ij} = \lambda \langle q\rangle ^{2\mu}
(q_iq_j)^{-\mu}$, we arrive again at (3). The exponent mapping
appears to be quite generally valid \cite{exceptions}.

\begin{figure}[h]
\includegraphics[width=2.5in,height=2in]{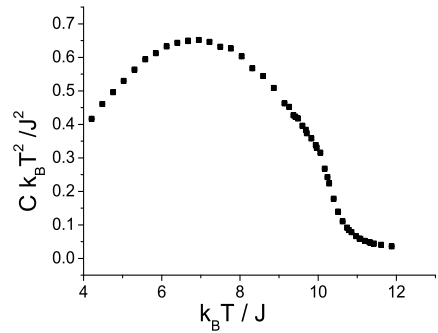}
\caption{\label{fig:epsart} Specific heat from MCS of BA networks
with $N=5600$, for $\mu = 1/3$ and $\langle q\rangle =10$. The
critical temperature in the large-$N$ limit is at $k_B T_c/J
\approx 11.33$ (Eq.6).}
\end{figure}

In conclusion, static and dynamic order-disorder transitions on
scale-free networks display singularities that depend on the
network topology {\em and} on the form of the interactions.
Connectivity-dependent interactions can be used as a probe of
topology-dependent cooperative behavior. The exponent mapping (3)
prescribes how to trade interactions for topology.

We thank Ton Coolen, Marcus M\"uller and Marco Baiesi
 for discussions and acknowledge grant FWO-G.0222.02.

\end{document}